\documentclass[a4paper, amsfonts, amssymb, amsmath, reprint, showkeys, nofootinbib]{revtex4-1}
\usepackage[english]{babel}
\usepackage[utf8]{inputenc}
\usepackage[colorinlistoftodos, color=green!40, prependcaption]{todonotes}

\usepackage[pdftex, pdftitle={Article}, pdfauthor={Author}]{hyperref}
\usepackage{float}
\usepackage{graphicx}
\usepackage{yhmath}
\usepackage{subfigure}
\usepackage{mathdots}
\usepackage{MnSymbol}

\begin{document}

\title{Complex matter field universal models with optimal scaling for solving combinatorial optimization problems}

	\author{ Natalia G. Berloff$^{1,2}$ }


	\email[correspondence address: ]{N.G.Berloff@damtp.cam.ac.uk}

	\affiliation{$^1$Department of Applied Mathematics and Theoretical Physics, University of Cambridge, Cambridge CB3 0WA, United Kingdom}
\affiliation{$^2$Skolkovo Institute of Science and Technology, Bolshoy Boulevard 30, bld.1, Moscow, 121205, Russian Federation}

\affiliation{Department of Applied Mathematics and Theoretical Physics, University of Cambridge, Cambridge CB3 0WA, United Kingdom}	
	\begin{abstract}{We develop a universal model based on the classical complex matter fields that allow the optimal mapping of many real-life NP-hard combinatorial optimisation problems into the problem of minimising a spin Hamiltonian. We explicitly formulate one-to-one mapping for three famous problems: graph colouring, the travelling salesman, and the modular N-queens problem. We show that such a formulation allows for several orders of magnitude improvement in the search for the global minimum compared to the standard Ising formulation. At the same time, the amplitude dynamics escape from the local minima.}
	\end{abstract}
	
	\maketitle
	{\it Introduction.} A vast majority of real-life optimisation problems are computationally impractical for conventional classical computers as the number of operations grows exponentially fast with the number of variables for such problems. Technological capabilities are hindered by the available computing resources required to solve such problems within relevant timescales subject to other constraints such a power consumption. Quantum computing is widely recognised as a potential solution; however, many technological challenges need to be overcome for this approach to become relevant for real-life problems such as mission control,
 dynamical analysis of financial markets, prediction of new chemical materials, machine learning, etc. 
 
 Technological demands, classical computing limitations and development of quantum computing hardware and algorithms have inspired the development of promising new algorithmic and novel unconventional analogue hardware techniques that, on the one hand, avoid the technical challenges facing quantum computing while, at the same time,  offering significant advantages over classical computing. The analog hardware based on   lasers \cite{lasers,nirprl2017,Babaeian2019}, optical parametric oscillators (OPOs) \cite{McMahon2016,Takata2016,Inagaki2016}, microring resonators \cite{Tezak2019}, polaritons \cite{BerloffNatMat2017,kalinin2020polaritonic,kalinin2018networks}, photonic systems \cite{Pierangeli2019,cen2020microwave}, photonic integrated circuits \cite{prabhu2019recurrent}, photon condensates \cite{leonetti2021optical}, confocal cavity QEDs \cite{marsh2020enhancing}, opto-electronic systems \cite{Chou2019,Boehm2019},  etc. emulates interacting dynamical systems that, at their steady-state, give solution to an optimization task. 
 
 The key to such opportunity lies in the {\bf universality} of a particular formulation. For instance, mathematically, it is possible to reformulate optimisation problems from vastly different areas into the problem of finding the ground state of a particular -- universal -- spin Hamiltonian with discrete or continuous degrees of freedom. The minimum of such a spin Hamiltonian can be found by a given simulator or by a special-purpose built analogue device, capable of incorporating the principle that drives the dynamical system to a high-quality solution of that optimisation task. 
 
 A particular and well-elucidated spin system is the Ising model on a graph that attracts most attention since a big range of hard discrete combinatorial optimization problems, e.g. travelling salesman, graph colouring, graph partitioning, etc. can be mapped into it with a polynomial overhead \cite{Lucas2014}. It is known that the Ising model  is universal \cite{DelasCuevas2016}  and many special-built computing devices (FPGAs, ASICs, etc.) are based on emulating classical Ising Hamiltonians \cite{toshiba,Tsukamoto2017,bohm2021not}. For instance, the Ising model can be minimised by the original Hopfield dynamical networks \cite{Hopfield1985} and many enhancements and modifications were proposed \cite{HopfieldModification1990, krotov2018dense, krotov2020large}. Many of the physical emulators of the Ising Hamiltonians have, at their core, the operation described by the Hopfield networks \cite{cai2020power,kalinin2022computational}.  
 
 The main obstacle in using the available purpose-built devices to solve real-life problems lies in driving to larger scales. Existing analogue hardware demonstrates that performance gains diminish exponentially as the number of interacting elements in the dynamical systems increases.
 When a combinatorial optimisation problem is embedded into the Ising Hamiltonian minimisation, the number of dynamical system elements typically grows at least as $O(N^2)$  where $N$ is the number of the original variables. The consequences of this increase, although polynomial, are vast. First, the quadratic increase in the number of variables increases the physical demands on the number of spins in the systems. Second, it requires increased analogue control precision for the increased number of elements in the dynamical systems. Third, it further drives the complexity of the optimisation landscape by increasing its dimensionality and roughness. Such inefficiency currently forces the corresponding circuit resources to grow according to a high-order power law in the problem size.
 
 It necessitates that we 
 develop the techniques where the dynamical system requirements grow (ideally) linearly with the number of problem variables. This is the question and challenge that this Letter addresses.
 
 To start, let's look at where the overhead occurs. It occurs when the states of the original variable cannot be represented as one binary number. For example, in the travelling salesman optimisation task, each city is associated with the order it is visited on the route. If one has $N$ cities, there are $N$ possible orders to visit each of the cities, giving an $N^2$ number of spins to represent the solution and $N^2$ number of dynamical equations to deal with. A vertex takes one of the possible $n$ colours for the graph colouring task. The N-queen problem requires $N^2$ spins to represent the chessboard squares.   

 The (at least) quadratic increase in the number of variables (and, therefore, in the dynamical equations or the dynamic elements in the system) is not the only problem. The analogue hardware has a lot of other limitations, such as those related to the topology of the dynamical systems. The type of interactions, the graph connectivity, how finely the interactions need to be controlled, and the range of the couplings play a significant role in the machine performance or for the class of problems that can be efficiently mapped to the analogue hardware. 
 In the Ising mapping of combinatorial problems, the constraints on the number of cities (for the travelling salesman problems), on the number of available colours (for the graph colouring problem), on the number of queens (in the $N$-queens problem) should come as a strict constraint. It means that the corresponding value of the coupling matrix has to be much larger than other coupling terms. Unfortunately, none of the current optimisers can deal with vastly varying coupling strengths.

 In this Letter, we propose a paradigmatic shift in mapping and problem representation for the analogue implementation of optimisation problems. It allows a linear (actually, one-to-one) mapping for the hard optimisation problems. It employs complex matter field representation leading to what we refer to as Complex matter Field Hamiltonian formulation (CFH).
 
 {\it A key function.} First of all,  we would like to switch to the complex domain for the elements of our system. We shall think of each element of the computing system ("bit") as an oscillator described by the complex field $\psi_i$ with two degrees of freedom: the amplitude, $a_i$, and the phase, $\theta_i$. This description lies at the core of the physical optical optimisers (OO) such as lasers, OPOs, polaritons, photons, etc. At the end of the search for the solution, the oscillator's state is represented by the phase $\theta_i$, while the amplitudes are used to modify (anneal) the target function. The Ising problem on a graph $G=(V,E)$ with vertices $V$ and edges $E$ with weights $W$ consists of finding the global minimum (GM) of the classical Ising Hamiltonian $H_I=-\sum_{i,j\ne i}W_{ij}s_i s_j,$ where $s_i=\{\pm 1\}.$ Quantum annealers such as D-Wave deal directly with $H_I$ as the target Hamiltonian. The specialised chip devices based on FPGAs or ASIC implement the matrix-vector multiplication and nonlinearity necessary to project continuous variables onto $\pm 1$ to represent  $H_I$ \cite{bohm2021not}.  OOs  aim at finding GM  of 
 \begin{equation}
     H_{G}=-\frac{1}{2}\sum_{i,j\ne i}W_{ij} (\psi_i\psi_j^*+c.c).
     \label{general}
 \end{equation}
If a degenerate excitation, second-order harmonic generation, or a resonant pumping limits the values of the phases to $\theta_i=0$ or $\pi$ and the amplitudes are brought to $a_i=1,$  then $H_G=H_I.$ If each phase is free to take any value in $[0,2\pi)$ (and $a_i=1$), then $H_G$ coincides with the classical XY Hamiltonian. In OO, the search for GM is based on a gain-dissipative dynamics \cite{kalinin2018global} driven by the principle of minimum power dissipation where the amplitude constraints can be included as the Lagrange multipliers \cite{vadlamani2020physics}, while the special purpose programmable chips may use threshold function for that. For instance, in the Coherent Ising machine, each Ising spin corresponds to a degenerate OPO that is described by a dynamics equation for the complex amplitude of the signal field $\psi_i$:
\begin{equation}
	\frac{d \psi_i}{dt} = \gamma(t) \psi_i^* - \psi_i - |\psi_i|^2 \psi_i -\frac{\delta H_G}{\delta \psi_i^*},
	\label{CIM_ai}
\end{equation}
where the dynamics are defined by an increasing gain $\gamma$ and   nonlinear  losses are normalised. For  polariton condensates or degenerate lasers, the dynamical equations are
\begin{equation}
	\frac{d \psi_i}{dt} = (\gamma(t) -|\psi_i|^2)\psi_i  - i U_0|\psi_i|^2 \psi_i -\frac{\delta H_G}{\delta \psi_i^*} + h(t) \psi_i^*,
	\label{polariton}
\end{equation}
 where $U_0$ is the strength of the self-interactions and $h(t)$ is the strength of the resonant excitation (if present). Eqs.~(\ref{CIM_ai}) and (\ref{polariton}) should be complemented by the processes that bring the amplitudes to one. The Hopfield networks realised by many physical platforms \cite{ramsauer2020hopfield} explicitly use phases $0$ and $\pi$, so the signal becomes real while amplitudes are projected into $\pm 1$ with dynamics that can be written as
 \begin{eqnarray}
     \frac{d Re[\psi_i]}{dt} &=& -\frac{\partial H_H}{\partial Re[\psi_i]},\nonumber\\ \quad H_H&=&  -\sum_{i,j\ne i} W_{ij} g_j -\gamma(t)\sum_{i=1}^N\int_0^{g_i}g^{-1}(x)\,dx,
	\label{Hopfield}
\end{eqnarray}
where e.g. $g(x)={\rm sgn}(x)$ or $g(x)=\tanh(lx),$ with some $l=l(t)$ increasing with time and $g_i=g(Re[\psi_i])$. The gradient descent used in these dynamical systems can be accelerated, for instance,  by momentum as in the Toshiba bifurcation FPGA \cite{toshiba} or by many other acceleration techniques. 
 
 So far, we have seen that the physical optimizer performs a variant of the gradient descent. However, the heterogeneity of the amplitudes allows the system to do more than just find the nearest local minima. During the time evolution, heterogeneity of the amplitudes allows "anneal" so to effectively tunnel between local minima in search of a lower-lying state or even for the actual GM. The details of the anneal schedules $\gamma(t), h(t), l(t),$ the structure of the projector $g(x)$ or the Lagrange multipliers for the amplitude's  constraint, as well as the scheme of the gradient descent all play an essential role in the performance of the solver. In this Letter, we propose a new set of basic operations and a universality structure associated with them, that dramatically reduces the number of the oscillators required to represent the original combinatorial problems while keeping the advantage offered by varying amplitudes in the search for the high-quality solution.
 
 We show the construction by  illustration. We start with the graph coloring problem: given $k$ different colors and a graph $G = (V,E)$,  how to color each vertex in the graph with a specific color, such that no edge connects two vertices of the same color? The Ising representation for this problem requires $kN$ spins and takes form $H_{GC}=\sum_{v\in V}(1 - \sum_{c=1}^kx_{vc})^2 + \sum_{(uv)\in E}\sum_{c=1}^k x_{uc}x_{vc},$ where $x_{vc}=(1+s_{vc})/2$ \cite{Lucas2014}. The spin $s_{vc}=1$ if vertex $v$ is colored with color $c$ and is $-1$ otherwise. The first term in $H_{GC}$ is the constraint (in the form of a penalty) on having a single color for each vertex, the second penalises if two vertices connected by an edge have the same color. Instead, we associate  a single oscillator $\psi_v$ with each vertex and encode the color associated with that vertex  into $k$ discrete values of the phase $\theta_v=\{0,2\pi/k,4\pi/k,...,2\pi(k-1)/k\}.$ The constraint on having a single color per each vertex is naturally satisfied with this representation. To express the second penalty we need to introduce a function that returns a positive value when $\psi_v$ and $\psi_u$ connected by an edge have the same phases (are colored the same) and is zero otherwise. Finally, there has to be a constraint leading to the discrete values of phases. We can take care of both conditions by observing that $\phi_k(\theta)\equiv |1 + \exp[i2\theta]+\exp[i3\theta]+...\exp[i (k-1)\theta]|^2=|(1-\exp[ik\theta])/(1-\exp[i \theta])|^2=\sin^2(k \theta/2)/\sin^2(
\theta/2)$, so that $\phi_k(0)=k^2$,  has the property that $\phi_k(i 2j\pi/k)=0,$ if integer  $j\ne 0,$ and takes positive values otherwise. The objective Hamiltonian replacing Eq.~(\ref{general}) for graph coloring problem becomes
\begin{equation}
    \widetilde{H_{GC}}=\sum_{(vu)\in E} \Phi_k(\psi_v,\psi_u),  \quad \Phi_k(\psi_v,\psi_u)\equiv\biggl|\frac{1-(\psi_v\psi_u^*)^k}{1-\psi_v\psi_u^*}\biggr|^2.
    \label{gc}
\end{equation}
 The functional $\Phi_k(\psi_u,\psi_v)$ defined in Eq.~(\ref{gc}) becomes the main building block of our CFH basis. 
 
 \begin{figure}  [t]
 \centering
\includegraphics[width=\columnwidth]{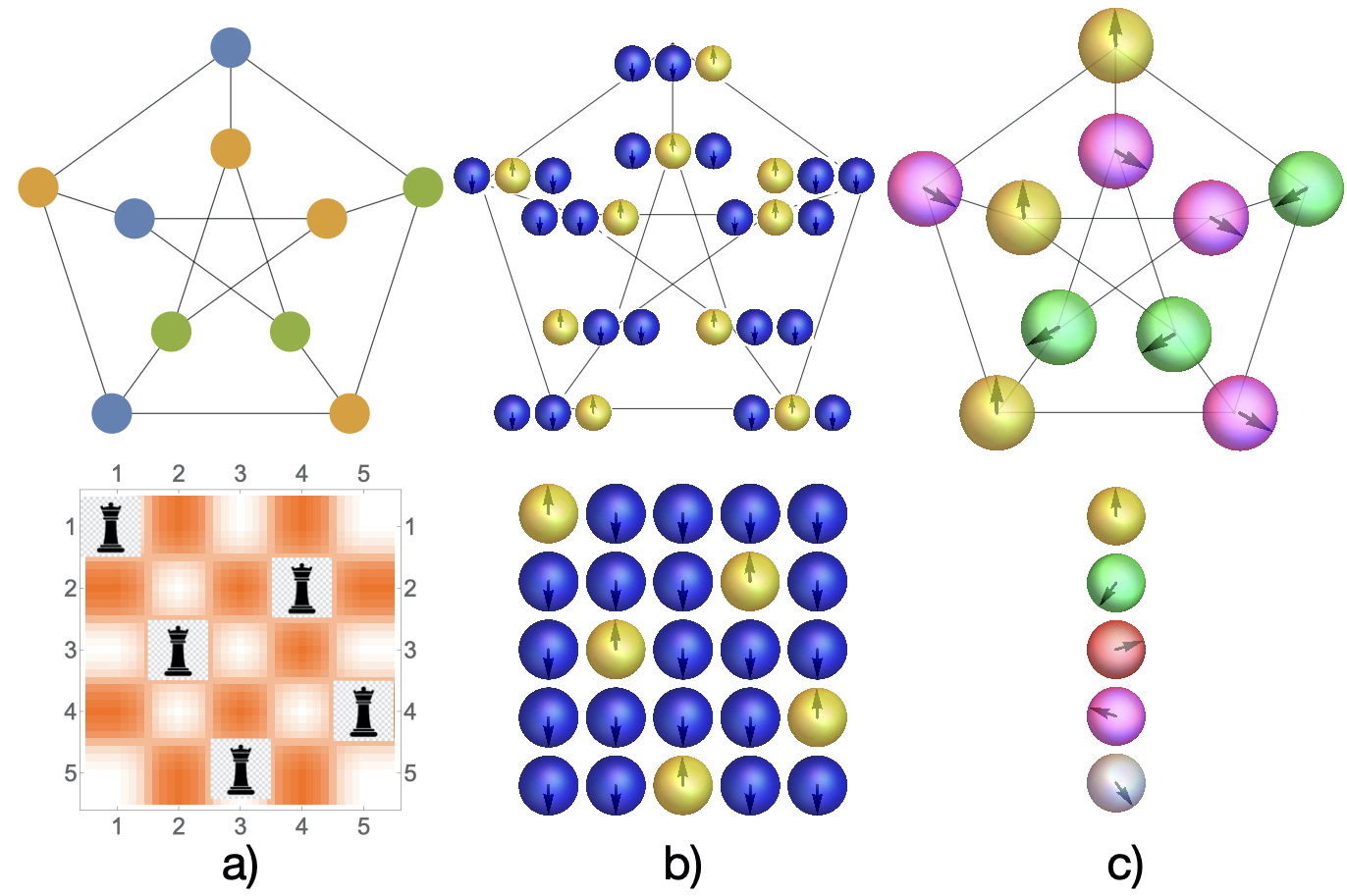}
\caption{Mapping for the two combinatorial problems discussed in the main text: graph colouring (top row) and the modular $N$-queens problem (bottom row). The solutions to the problems are shown in (a). (b) depicts the Ising formulation of these problems where each vertex of the graph colouring problem is mapped into three Ising spins, and each board square of the N-queens problem is mapped into one Ising spin. (c) depicts the CFH formulation where each vertex colour is mapped into one discrete phase $\{0, 2\pi/3, 4\pi/3\}$ for the graph colouring problem. The position of each queen is mapped into a single discrete phase $\{0,2\pi/5,4\pi/5,6\pi/5,8\pi/5\}.$  }
\label{schematics}    
\end{figure} 
\begin{figure}[t!]
 \centering
\includegraphics[width=0.8\columnwidth]{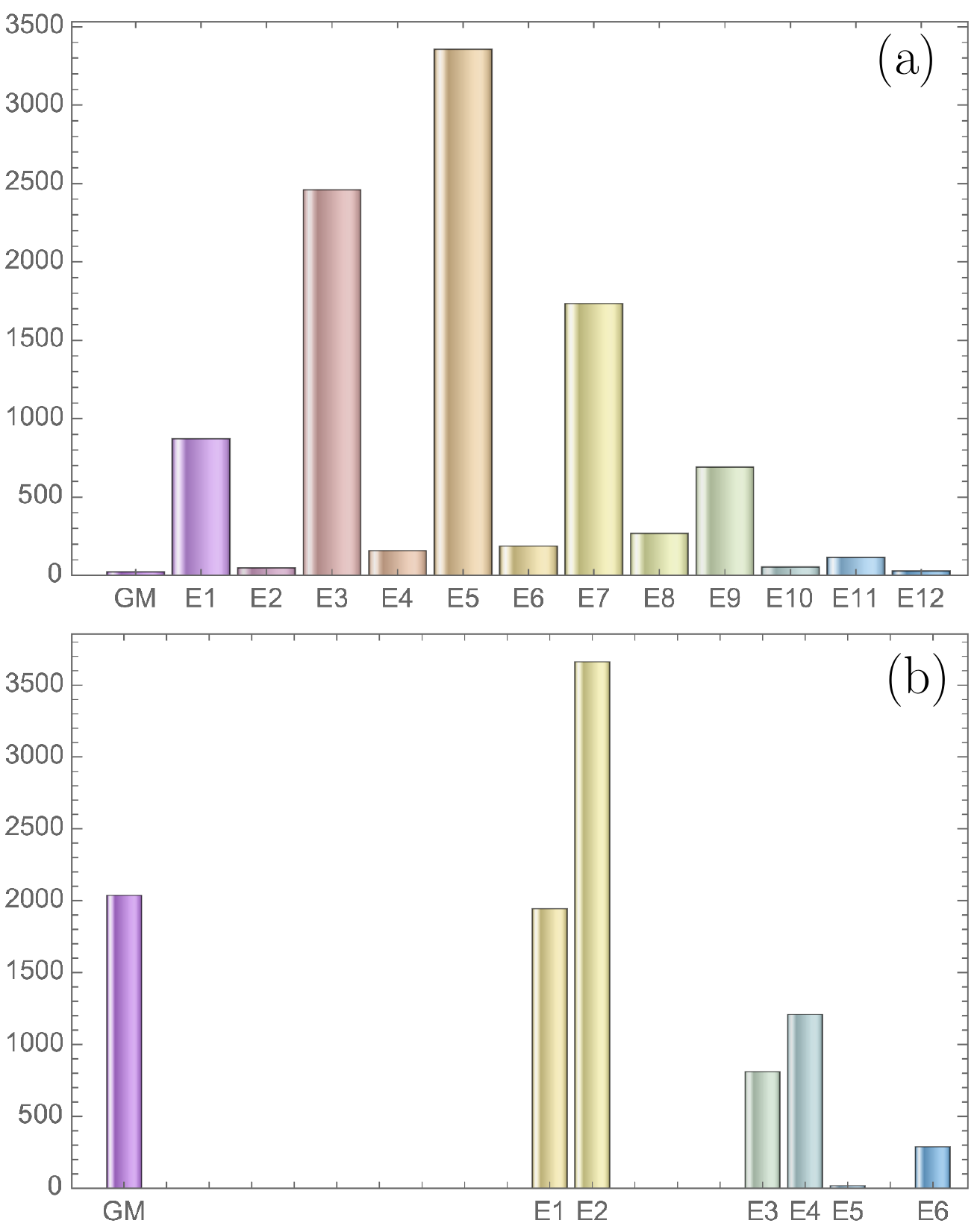}
\caption{The structures of the Hamiltonian in different representations of the modular $N$-queens problem. The  Monte Carlo simulations assisted by the Broyden–Fletcher–Goldfarb–Shanno quasi-Newton method are used to elucidate the structure of the  Ising (Eq.~(\ref{queens}) (a) and  the   CFH (Eq.~(\ref{myqueens}) (b) Hamiltonians for $N=5$ and $a_i=1$. We use $10,000$ random initial conditions  to find the minimum. The statistics of the minima found is given for the Ising (a) and CFH (b) formulations. The success probability of finding the GM is more than $1/5$ for the CFH, whereas the Ising formulation of the same problem is two orders of magnitude lower with $0.0023$. The CFH also has a large spectral gap separating the GM from the first excited state. For the Ising formulation we used its continuous representation with $x_{ij}=(\cos(\theta_{ij})+1)/2$  and the penalty $M\sum_{ij} \cos(2 \theta_{ij})$ to project phases onto $0$ and $\pi$ added to $H_Q.$ The optimal parameters for minimisation were found and used as $B=2$, $M=1$. }
\label{statistics}    
\end{figure} 

 Travelling salesman problem is another famous NP-hard problem that searches for the Hamiltonian cycle in the graph $G=(V,E),$ $|V|=N$, such that the sum of the weights, $w_{uv}$ of each edge in the cycle is minimized. The Ising formulation of this problem is \cite{Lucas2014} 
 \begin{eqnarray}
     H_{TS}&=&A\sum_{v=1}^N (1 - \sum_{i=1}^N x_{vi})^2 +A\sum_{i=1}^N (1 - \sum_{v=1}^N x_{vi})^2 \nonumber \\
     &+&  \sum_{u,v\ne u}W_{uv}\sum_{i=1}^N x_{ui}x_{v i+1},
     \label{ts}
 \end{eqnarray}
 where $x_{vi}=1$ if the vertex labeled $v$ is visited in the step  $i$. We assumed a cycle notation $N+1=1$ and introduced $W_{uv}=w_{uv}$ if $(uv)\in E$ and $W_{uv}={\bar w}>\max(w_{uv})$ if $(uv)\nin E$. Total $N^2$ spins are required for the mapping into the Ising formulation.  To fulfil the constraints (each vertex and each order appear exactly once), we have to take $A\gg {\bar w},$ which leads to a large spread of the coefficients that hinders minimisation success.
 
 Instead, we associate one oscillator $\psi_v$ with each vertex (requires $N$ oscillators). The objective to minimise becomes
 \begin{equation}
   \widetilde{H_{TS}}= \sum_{u,v \ne u}W_{uv} \Phi_N(\psi_u,\psi_v \times e^{i \frac{2\pi}{N}}).
   \label{myts}
 \end{equation}
 
 In the last example,
 we consider the $N-$queens problem that tasks placing $N$ queens on an $N$x$N$ chessboard in non-attacking positions (queens attack along rows, columns, and diagonals). The variations of this problem (with blocked diagonals or some queens already placed on the board) is an NP-hard problem \cite{gent2017complexity}. It is a straightforward exercise to introduce such variations into the formulation, so we will not consider them here for brevity. We also formulate our representation for the queen's placement on the doubly periodic board (so-called modular $N$-queens problem). The Ising formulation requires $N^2$ spins: $s_{ij}=1$  if there is a queen in the $i-$th row and $j-$th column and $-1$ otherwise. The Hamiltonian in the Ising formulation \cite{torggler2019quantum} reads
 \begin{equation}
   H_Q=B(N-\sum_{i=1}^N \sum_{j=1}^N x_{ij})^2 +  +\sum_{ijkl=1}^N W_{ijkl} x_{ij}x_{kl},  
   \label{queens}
 \end{equation}
where the weights are $W_{ijkl}=1$ if the squares $(ij)$ and $(kl)$ are on the same diagonal, row or column and zero otherwise. The first term is the constraint on the number of queens with $B\gg 1$. With the growth in $N$, coupling strengths in the matrix of interactions grow as $BN$, leading to a vast difference in coupling strengths between the terms. 

In CFH representation, we associate one oscillator $\psi_k$ with the queen in the $k-$th row, so the constraint on the number of queens is naturally satisfied and the Hamiltonian becomes
\begin{eqnarray}
 \widetilde{H_{Q}}&=&\sum_{j=1,k>j}^N\Phi_N(\psi_k, \psi_j) +\sum_{j=1,k\ne j}^N\Phi_N(\psi_k e^{i\frac{2\pi(k-j)}{N}},\psi_j)\nonumber\\&+&\Phi_N(\psi_k,\psi_j e^{i\frac{2\pi(k-j)}{N}}). 
 \label{myqueens}
\end{eqnarray}
 Figure \ref{schematics} illustrates  the differences in Ising and CFH representations for the graph coloring problem Fig.~(\ref{schematics})(top row) and the modular N-queens problem Fig.~(\ref{schematics})(bottom row). 
 
 \begin{figure}[t!]
\includegraphics[width=0.8\columnwidth]{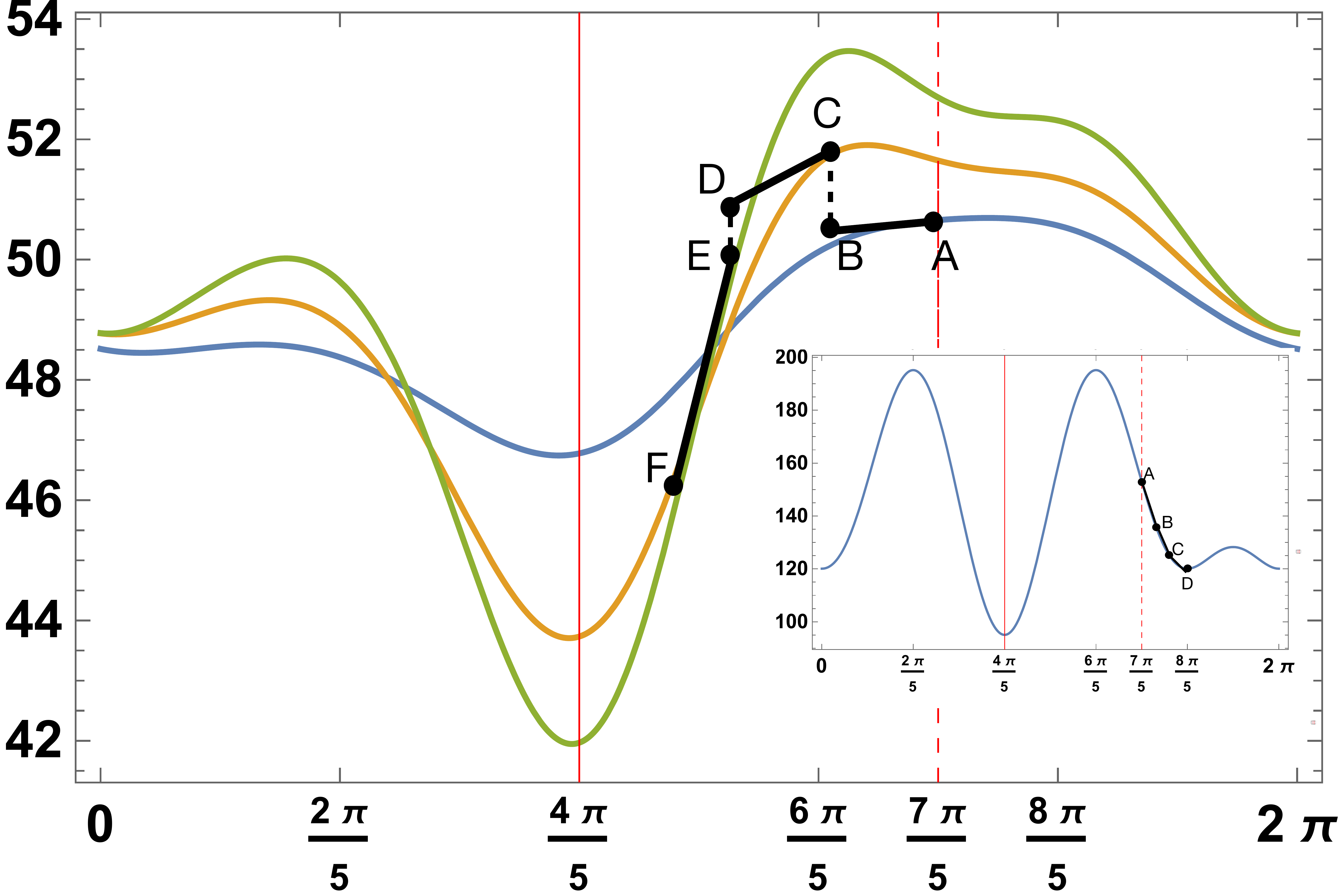}
\caption{Time evolution of the target landscape with amplitudes assisting the search for the GM. The inset shows the projection of the target Hamiltonian $\widetilde{H_Q}$ of Eq.~(\ref{myqueens}) with $a_i=1$ for $N=5$ to the plane with $\theta_1=0, \theta_2=6\pi/5, \theta_3=2\pi/5, \theta_4=8 \pi/5$, so only $\theta_5$ remains to be found. 
The optimal value is at $\theta_5=4\pi/5$, however, if the initial point is chosen as $\theta_5=7\pi/5$ (marked "A" in the inset), the gradient descent ($A\rightarrow B\rightarrow C \rightarrow D$) brings it to the excited state with $\theta_5=7\pi/5$.
However, the amplitudes change the structure of the function to be minimised. On the main graph we show $\widetilde{H_Q}$ of Eq.~(\ref{myqueens}) with $\theta_1=0, \theta_2=6\pi/5, \theta_3=2\pi/5, \theta_4=8 \pi/5$ but this time the amplitudes are not $1$, instead they are changing gradually. 
Starting with the same point as in the inset
$\theta_5=7\pi/5$, the gradient decent brings the solution to the true GM. The graphs shown
corresponds to $a_1=1, a_2=0.5, a_3=0.9, a_4=0.4, a_5=0.4$ (blue), $a_1=1, a_2=0.7, a_3=0.9, a_4=0.4, a_5=0.55$ (orange), $a_1=1, a_2=0.75, a_3=0.95, a_4=0.45, a_5=0.6$ (green). The gradient decent follows $A\rightarrow B\rightarrow C \rightarrow D \rightarrow E \rightarrow F$ route to the GM.
 }
\label{gd}    
\end{figure} 
 
 When the original problem is mapped into a universal Hamiltonian, the dynamical system can be tuned to find the solution, so Eqs.~(\ref{CIM_ai}, \ref{polariton}, \ref{Hopfield}) etc.  can be employed with the CFH formulations reflected by  $\widetilde{H_{GC}}, \widetilde{H_{TS}}, \widetilde{H_{Q}}$ replacing the Ising Hamiltonians $H_G$ or $H_H$. The effectiveness of the search for the solution depends on the structure and parameters of the dynamical system employed, however, it is correlated with the structure of the target Hamiltonian.  We compare the target Ising Hamiltonians given by Eq.~(\ref{queens}) with the CFH defined by  Eq.~(\ref{myqueens}) with $|\psi_k|^2=1$ for all $k$ using the Monte Carlo simulations. Figure (\ref{statistics}) summarizes our findings showing that CFH formulation brings two orders of magnitude improvement over the Ising formulation even for  small to moderate tasks, while this advantage is bound to grow exponentially for larger tasks. Figure (\ref{gd}) illustrates the importance of the amplitude dynamics in finding the GM.  
 
 The mapping of $N$ states into the discrete values of the phase becomes problematic as $N$ grows. There is a limit on the precision with which the phases can be measured for the physical implementation. Also, the discrepancy between the values of $\Phi$ at zero and on non-discrete phase values increases with $N$, hindering the algorithmic performance. This problem can be overcome by associating two or more oscillators with each vertex and mapping $N$ states of the vertex into the discrete phases of oscillators representing each state. For instance, let's assume that only $p$ discrete values of the phase can be experimentally resolved. In this case, we can encode $N$ states into $\lfloor \log_p N\rfloor$ oscillators.  When the Ising formulation  requires $N^2$ spins (and $N^2$ dynamical equation in OO), the CFH  requires only $\lfloor \log_p N\rfloor N$ oscillators and dynamical equations. The  effect of this modification on the Hamiltonian structure and performance will be considered elsewhere.

 {\it Discussion.}We proposed a complex field Hamiltonian mapping that allows overcoming the power-law scaling of the Ising Hamiltonian mapping that is typical for real-life problems. Such CFHs are universal (any NP problem can be mapped into them) and consist of a well defined set of the operations described by the function $\Phi_k(\psi_i,\psi_j)$ of Eq.~(\ref{gc}) and  may require the implementation of the phase shifts $\psi\rightarrow\psi \exp[i \theta]$ for some $\theta$. The dramatic reduction in the number of variables is accompanied by an improved structure of the Hamiltonian suitable to various OO dynamical implementations. The annealing of the amplitudes allows the transition away from the local minima in the search for a high quality (low energy) solution.

\bibliography{references,refs,Refs-pr2}{}
\bibliographystyle{ieeetr}

\end{document}